\documentclass[a4paper,twocolumn,11pt,accepted=2022-12-21]{quantumarticle}

\pdfoutput=1
\usepackage[utf8]{inputenc}
\usepackage[english]{babel}
\usepackage[T1]{fontenc}
\usepackage{bm,bbold}
\usepackage{bbm}
\usepackage{amsfonts, amsmath, amsthm, amssymb} 
\usepackage{physics}
\usepackage{hyperref}
\usepackage[numbers,sort&compress]{natbib}

\begin{document}

\title{Probing quantum entanglement from magnetic-sublevels populations: beyond spin squeezing inequalities}

\author{Guillem M\"uller-Rigat}
\email{guillem.muller@icfo.eu}
\affiliation{ICFO - Institut de Ciencies Fotoniques, The Barcelona Institute of Science and Technology, 08860 Castelldefels (Barcelona), Spain}

\author{Maciej Lewenstein}
\affiliation{ICFO - Institut de Ciencies Fotoniques, The Barcelona Institute of Science and Technology, 08860 Castelldefels (Barcelona), Spain}
\affiliation{ICREA, Pg. Llu\'{\i}s Companys 23, 08010 Barcelona, Spain}

\author{Ir\'en\'ee Fr\'erot}
\email{irenee.frerot@neel.cnrs.fr}
\affiliation{Univ Grenoble  Alpes, CNRS, Grenoble INP, Institut Néel, 38000 Grenoble, France}
\orcid{0000-0002-7703-8539}

\begin{abstract}
 Spin squeezing inequalities (SSI) represent a major tool to probe quantum entanglement among a collection of few-level atoms, and are based on collective spin measurements and their fluctuations. Yet, for atomic ensembles of spin-$j$ atoms and ultracold spinor gases, many experiments can image the populations in all Zeeman sublevels $s=-j, -j+1, \dots, j$, potentially revealing finer features of quantum entanglement not captured by SSI. Here we present a systematic approach which exploits Zeeman-sublevel population measurements in order to construct novel entanglement criteria, and illustrate our approach on ground states of spin-1 and spin-2 Bose-Einstein condensates. Beyond these specific examples, our approach allows one to infer, in a systematic manner, the optimal permutationally-invariant entanglement witness for any given set of collective measurements in an ensemble of $d$-level quantum systems.
\end{abstract}

\maketitle

\section{Introduction}
To prepare and detect quantum-entangled states has become a central goal for experimental many-body physics. On the one hand, it demonstrates the ability to probe regimes unthinkable within a classical framework, either at equilibrium or during the dynamics. On the other hand, it represents a crucial step towards using such systems as resources for quantum-enhanced applications, such as sensing \cite{pezzeetal2018,gilmoreetal2021}, or the simulation of quantum many-body problems \cite{georgescuetal2014}.

Here, our main focus are those many-body systems which are most easily probed by collective measurements, such as atomic ensembles and ultracold spinor gases. In such systems, entanglement can be famously revealed by so-called spin-squeezing inequalities (SSI), which involve measurements of first- and second-moments of collective spin operators \cite{kitagawa_ueda_1993,winelandetal1994,sorensenM2001,sorensenetal2001,tothetal2009,vitaglianoetal2011,Ma_2011,vitaglianoetal2014,pezzeetal2018}. SSI and their generalizations \cite{tothetal2009,vitaglianoetal2011,vitaglianoetal2014,gessneretal2019} have found an impressively broad range of successful applications, from detecting entanglement in cold and hot atomic ensembles \cite{pezzeetal2018,Kongetal2020} to unveiling many-body Bell non-locality \cite{schmiedetal2016,engelsen_bell_2017,PhysRevLett.123.170604,PRXQuantum.2.030329}, from probing quantum phase transitions \cite{PhysRevLett.121.020402} and quantum quenches \cite{PhysRevA.105.022625,Comparin_2022} to applications in quantum-enhanced metrology \cite{pezzeetal2018}. In particular, a handful of generalized SSI is sufficient to capture the complete palette of entanglement that can be revealed via collective-spin measurement in $j=1/2$ spin ensembles \cite{tothetal2009}. In the context of $j=1$ spinor Bose-Einstein condensates (BEC), by restricting the spin states to effective two-dimensional subspaces, entanglement has also been probed via those SSI \cite{duanetal2002,hamleyetal2012,kunkeletal2018,kunkeletal2019,quetal2020}. In parallel, the last decade has witnessed the development of several experiments investigating much-larger-$j$ spinor gases \cite{frischetal2014,quetal2020,evrardetal2021,evrardetal_science2021,hamleyetal2012,chomazetal2016,wenzeletal,kadauetal2016,luoetal2017,zouetal2018,kunkeletal2018,lucionietal2018,langeetal2018,kunkeletal2019,trautmannetal2018,Chalopin_2018,batailleetal2020,gabardosetal2020}, mixtures \cite{kasper2021universal,bhattetal2022}, or effective qudits ensembles \cite{ringbauer2021universal}. In these systems, the populations of all Zeeman sublevels $s=-j, -j+1, \dots, j$ can be measured, typically by fluorescence imaging after collective spin rotations and Stern-Gerlach splitting \cite{hamleyetal2012,zouetal2018,quetal2020,evrardetal2021,evrardetal_science2021,luoetal2017,kunkeletal2018,kunkeletal2019,langeetal2018,gabardosetal2020}. The SSI generalized in refs.~\cite{vitaglianoetal2014} to spin $j>1/2$ can be potentially be applied in such systems in order to detect entanglement. Yet, the entanglement patterns incorporated by $j>1/2$ spin ensembles is potentially much richer than what can be captured by SSI, with new classes of entangled states and new entanglement mechanisms. The optimal use of such collective imaging data to probe entanglement beyond SSI has remained an open problem, for which our paper offers a novel approach.

In order to do so, we actually formulate and solve a more general problem: considering a set of single-atom observables, we assume that their average value and pair correlations, averaged over all permutations of the atoms, can be estimated. In ref.~\cite{vitaglianoetal2011}, a similar framework was proposed, and new entanglement witnesses were derived. Yet, in order to potentially find one violated inequality, in general the latter framework requires to test a number of inequalities which grows exponentially with the number of local observables, such as populations in different Zeeman sublevels as we consider here. In this work, we instead derive a single inequality based on such data to reveal entanglement. This inequality not only summarizes all known SSI \cite{tothetal2009,vitaglianoetal2011,vitaglianoetal2014}, but also detects new forms of entanglement. Specializing then to Zeeman-sublevels population measurements in spinor gases, we apply this general approach to discover novel families of entanglement criteria, akin to but not captured by SSI. We illustrate our findings on representative states of a spin-1 BEC, as produced and characterized in several experiments \cite{hamleyetal2012,zouetal2018,quetal2020,evrardetal2021,evrardetal_science2021}. Finally, we extend these findings to the largely unexplored case of a spin-2 BEC.

\paragraph{Framework.}
Our goal is to exclude the possibility to decompose $\hat \rho$, the many-body density operator of the system, as a statistical mixture of product states over individual atoms: $\hat \rho \neq \hat \rho_{\rm sep} := \sum_\lambda p_\lambda \otimes_{i=1}^N \hat \rho_\lambda(i)$. Here, $\hat \rho_\lambda(i)$ is an arbitrary internal state (pure or mixed) for atom $i=1,\dots,N$, and $p_\lambda>0$ is an arbitrary probability distribution, defining a separable state $\hat \rho_{\rm sep}$.  Our starting point is to consider for each atom a set of $K$ different observables $\{\hat o_k(i)\}_{k=1}^K$, some of which must be non-commuting in order to potentially detect entanglement. We then assume that data of the following form are available:
\begin{align}
	& m_k = \frac{1}{N} \sum_{i=1}^N \langle \hat o_k(i) \rangle ~, 	\label{eq_Qdata_1body}
\\
	& C_{kl} = \frac{1}{N(N-1)} \sum_{i \neq j} \langle \hat o_k(i) \hat o_l(j) \rangle ~,
	\label{eq_Qdata_2body}
 \end{align}
 where $\langle \dots \rangle = {\rm Tr}(\hat \rho \dots)$ is an expectation value over many identically-prepared systems. In practice, as we illustrate below, the correlations $C_{kl}$ might be available for only a subset of pairs $(k,l)$, something which can be accommodated within our approach. It is also central to our method to find the bound $\beta$ defined as:
\begin{equation}
	\beta = \frac{1}{N} \sum_{i=1}^N \max_{|\psi\rangle} \left\{ \sum_{k=1}^K |\langle \psi| \hat o_k(i)|\psi \rangle|^2\right\} ~,
	\label{eq_betaBound}
\end{equation}
where $|\psi\rangle$ is an arbitrary internal state of atom $i$. Depending on the choice of observables $\hat o_k(i)$, the bound $\beta$ might be computable analytically, but in general it may be found numerically, optimizing over the states of atom $i$. In the examples studied in this work, the local observables $\hat o_k(i)$ are the same for all atoms (they are the components of the spin, or projectors onto spin states), and we simply parametrize $|\psi\rangle=\sum_{\alpha=1}^d c_\alpha|\alpha\rangle$ with its complex coefficients $c_\alpha$ in a fixed basis. $\sum_{k=1}^K|\langle \psi| \hat o_k(i)|\psi \rangle|^2$ is then a function of the coefficient $c_\alpha$ (and is independent of $i$ in the examples we study), that we optimized using standard numerical routines. 

Our central result is the following inequality, valid for any separable state of the many-body system:
 \begin{equation}
	 N ~{\rm Tr}[ P( C - {\bf m} \otimes {\bf m})]  \ge {\rm Tr}(C)  - \beta ~,
	 \label{eq_central_ineq}
\end{equation}
where we introduced the notation $[{\bf m} \otimes {\bf m}]_{kl} = m_k m_l$, and $P$ is any projector (namely a $K \times K$ symmetric matrix obeying $P^2=P$). The detailed proof of Eq.~\eqref{eq_central_ineq} is given in Appendix \ref{app_main_result}. The optimal choice to most strongly violate Eq.~\eqref{eq_central_ineq} is to project onto the subspace corresponding to all negative eigenvalues of the matrix $C - {\bf m}\otimes {\bf m}$. As a special case, Eq.~\eqref{eq_central_ineq} contains all generalized SSI of ref.~\cite{vitaglianoetal2014} when choosing the spin in orthogonal directions $x,y,z$ as local observables (see Appendix \ref{app_SSI} for a detailed derivation). As we shall see, when applied to suitable states of a spinor gas, it allows us to find novel entanglement criteria beyond SSI.

\paragraph{Zeeman-sublevels population measurements.}
In order to illustrate the relevance of our new approach to probe entanglement in a spin-$j$ atomic ensemble, we assume that the populations $\hat N_{a,s}$ of all Zeeman sublevels $s=-j,\dots,j$ can be measured, for different quantization axes $a$ (at least, for the purpose of entanglement detection, along two different orientations). This is typically achieved by first applying collective spin rotations to map a given orientation $a$ onto the $z$ axis, followed by a spatial splitting of the atoms in a magnetic-field gradient. The intensity of fluorescence imaging finally allows one to estimate the number of atoms in each Zeeman sublevel. For each orientation $a$, average populations $\langle \hat N_{a,s} \rangle$ and their correlations $\langle \hat N_{a,s} \hat N_{a,s'} \rangle$ can be inferred in this manner. In the general framework presented above, the single-atom observables correspond to the projectors onto the spin states $\hat{n}_{a,s}(i)= |s_a\rangle_i \langle s_a|_i$ (namely, the label $k$ of previous section is now a pair of labels $(a,s)$ denoting both the quantization axis and the spin state). Correspondingly, one reconstructs the average values $m_{a,s} = \langle \hat N_{a,s} \rangle / N$, and the diagonal blocks of the correlation matrix corresponding to a fixed orientation $C_{(a,s)(a,s')} = \langle \hat N_{a,s} [\hat N_{a,s'} - \delta_{s,s'}] \rangle / [N(N-1)]$. The projector $P$ in Eq.~\eqref{eq_central_ineq} is then chosen in a block-diagonal form, projecting onto the negative eigenspace of each block, defined by the matrices $C_{(a,s)(a,s')} - m_{a,s} m_{a,s'}$ for the different quantization axes $a$.\\

\section{Applications to a spin-1 BEC} We first illustrate our approach on a BEC of spin-1 atoms, as investigated recently in several experiments  \cite{hamleyetal2012,zouetal2018,quetal2020,evrardetal2021,evrardetal_science2021}. We consider $N$ spin-$1$ atoms with contact isotropic interactions in the single-mode approximation, adiabatically prepared from the so-called polar state $|0,N,0\rangle=\otimes_{i=1}^N|0_z\rangle_i$ (with $|N_-, N_0, N_+\rangle$ the state written in the collective population basis along the $z$ quantization axis). In an external magnetic field, the Hamiltonian reads (see Appendix \ref{app_spinor_model} for details):
\begin{equation}
    \hat{H}_{j=1}=\frac{c}{N}\hat{\mathbf{J}}^2 + q\hat{Q}_z ~,
    \label{H_spin-1}
\end{equation}
where $\hat J_a = \sum_{s=-j}^j s \hat{N}_{a,s}$ is the collective spin in direction $a \in \{x,y,z\}$ and $\hat{\mathbf{J}}^2 = \hat J_x^2 + \hat J_y^2 + \hat J_z^2$; $\hat{Q}_z = \sum_{s=-j}^j s^2 \hat{N}_{z,s}$ is the collective quadrupole operator; and $\{x,y,z\}$ is an orthornormal basis of $\mathbb{R}^3$. Notice that the linear Zeeman term, proportional to $\hat{J}_z$, is omitted as it commutes with $\hat{\mathbf{J}}^2$ and acts trivially on the initial state: $\hat{J}_z|0,N,0\rangle = 0$. The initial state is also the ground state of $\hat{Q}_z$; and by varying the magnetic field intensity ($q$) one can adiabatically prepare the ground state of $\hat{H}_{j=1}$ for either ferromagnetic interactions ($c \le 0$) or antiferromagnetic interactions ($c \ge 0$). Notice that in the case of ferromagnetic interactions, two quantum phase transitions occur at $q/|c|=\pm 4$, so that adiabaticity is possible only for finite $N$. As the Zeeman sublevel populations are related by $\sum_{s=-j}^j\hat{N}_{a,}=N$ for all directions, $m_{a,s}$ and $C_{(a,s)(a,s')}$ for all $s,s'$ are not all independent. As local observables in Eqs.~\eqref{eq_Qdata_1body} and \eqref{eq_Qdata_2body} we used the spin $\hat S_a(i)=\sum_{s=-j}^j s \hat{n}_{a,s}(i)$ and the projector onto the Zeeman sublevel $s=0$, $\hat{n}_{a,0}(i)$. We use these data as input to Eq.~\eqref{eq_central_ineq}, with bound $\beta_{j=1}=3/2$ in Eq.~\eqref{eq_betaBound} (see Appendix \ref{app_EW}). Notice that for $j=1$, we have $\hat{Q}_a = \hat N_{a,+} + \hat N_{a,-} = N - \hat N_{a,0}$. \\

\begin{figure}[ht!]
	\includegraphics[width=\linewidth]{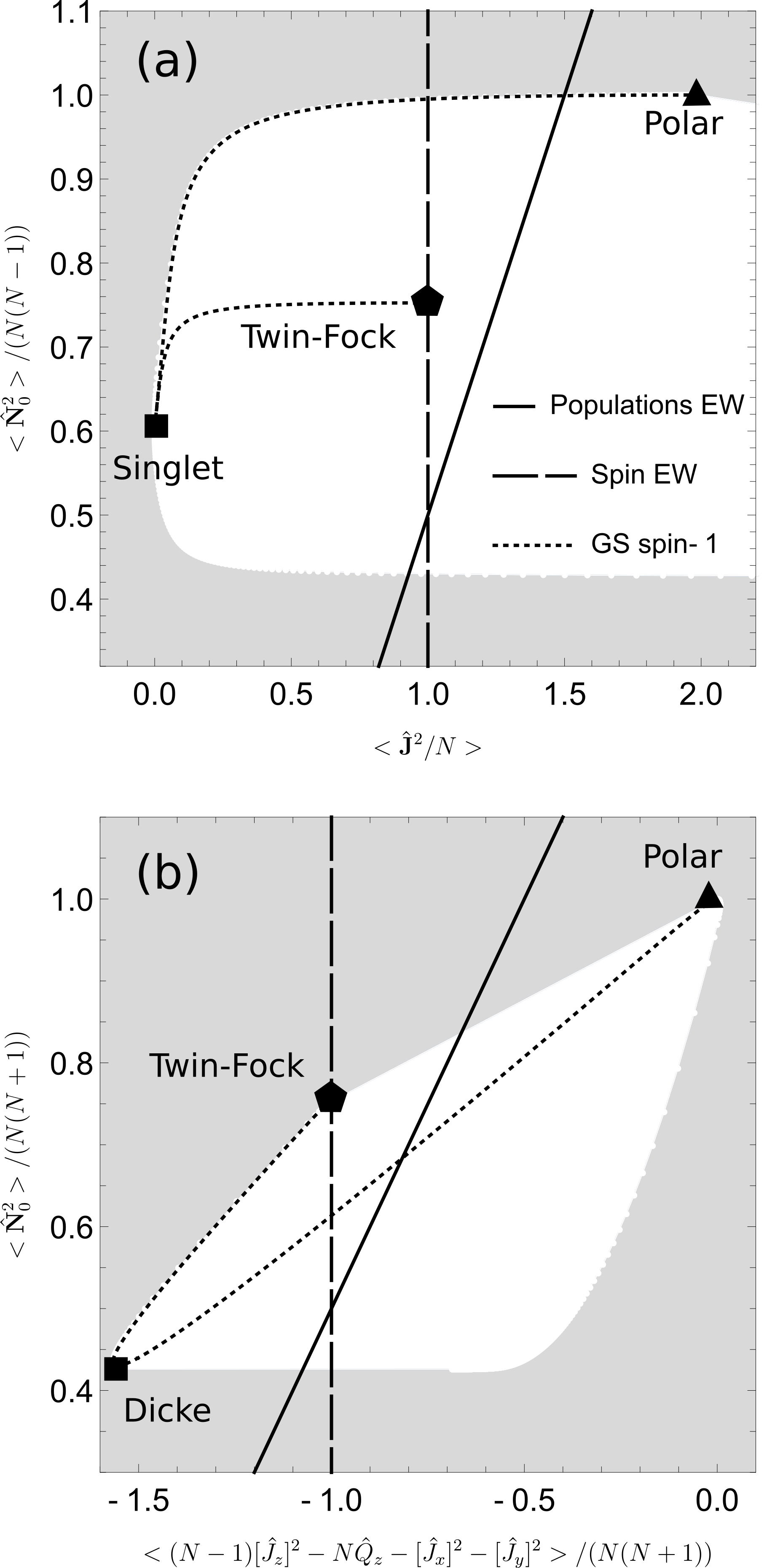}
	\caption{(a) Comparison of the population-based witness [Eq.~\eqref{EW_singlet_spin-1}] (solid line) with the SSI ${\rm Var}(\hat{\bf J})\ge Nj$ (dashed line) for spin-1 states ($N=100$ atoms). The light grey area are data points which cannot be achieved (see text and Appendix \ref{app_feasible}). The dotted line indicates the ground state of Eq.~\eqref{H_spin-1} for both antiferromagnetic interactions ($c=1$), varying the quadratic Zeeman term ($-10\le q \le 10$). Three representative states are also indicated: the polar state $|0,N,0\rangle$ ($q\to \infty$) which is separable, the spin singlet s.t. $ \hat{\bf J}^2 |\Psi\rangle=0$ ($q=0$), and the twin-Fock state $|N/2,0,N/2\rangle$ ($q\to -\infty$). (b) Comparison of the population-based witness [Eq.~\eqref{EW_z_spin-1}]  with a related SSI (see text) for spin-1 states. Graphical conventions as in panel (a). Here, the ground state is computed for ferromagnetic interactions ($c=-1$). At $q=0$, we obtain the Dicke state $\hat{\bf J}^2 |{\rm Dicke}\rangle=N(N+1)|{\rm Dicke}\rangle$, $\hat J_z|\Psi\rangle=0$.}
	\label{fig_spin_1}
\end{figure}

\paragraph{Antiferromagnetic interactions.}
For antiferromagnetic interactions ($c>0$) and $q/c \lesssim 4.4 $ ($N=100$), we find violation of Eq.~\eqref{eq_central_ineq}, with an entanglement witness (EW) which reads (see Appendix \ref{app_EW}):
\begin{equation}
   \mathrm{Var}(\hat{\mathbf{J}})-\frac{1}{N-1}\langle \hat{\mathbf{N}}_0^2\rangle  \geq \frac{N(N-3)}{2(N-1)}\ ,
    \label{EW_singlet_spin-1} 
\end{equation}
with the notations $ \mathrm{Var}(\hat{\mathbf{J}})=\mathrm{Var}(\hat{J}_x)+\mathrm{Var}(\hat{J}_y)+\mathrm{Var}(\hat{J}_z)$, $\mathrm{Var}(\hat{O})=\langle\hat{O}^2 \rangle-\langle\hat{O} \rangle^2$ and $\hat{\mathbf{N}}_0=(\hat{N}_{x,0},\hat{N}_{y,0},\hat{N}_{z,0} )$. The EW of Eq.~\eqref{EW_singlet_spin-1} is maximally violated at $q=0$, where the ground state of Eq.~\eqref{H_spin-1} is a spin singlet ($\mathrm{Var}(\hat{\mathbf{J}})=0$). In particular, we also applied our method using as input the experimental data of ref.~\cite{evrardetal_science2021}, where an almost perfect spin singlet is prepared, and all populations are measured; we also recovered Eq.~\eqref{EW_singlet_spin-1} as the optimal violated EW. This EW is reminiscent of, yet generally more powerful than, the generalized SSI $\mathrm{Var}(\hat{\mathbf{J}}) \ge Nj$ (with $j=1$ here) derived in ref.~\cite{vitaglianoetal2014}. For a quantitative comparison of both criteria, in Fig.~\ref{fig_spin_1}(a) we illustrate their detection power for $N=100$ and states with $\langle \hat{\mathbf{J}} \rangle=\mathbf{0}$ (vanishing mean spin, as is the case in the ground state of $\hat{H}_{j=1}$, Eq.~\eqref{H_spin-1}) in the plane $(x=\langle \hat{\mathbf{J}}^2 \rangle / N, y=\langle \hat{\mathbf{N}}_0^2\rangle/[N(N-1)])$, where violation occurs respectively for $x-y<(N-3)/[2(N-1)]$  [Eq.~\eqref{EW_singlet_spin-1}] and $x<1$ (SSI). We plot the data points corresponding to the ground state of $\hat{H}_{j=1}$ for $c>0$ and both $q>0$ and $q<0$. More generally, not all combinations of $\langle \hat{\bf J}^2 \rangle$ and $\langle \hat{\bf N}_0^2 \rangle$ are physically allowed. As detailed in Appendix \ref{app_feasible} (see also \cite{sorensen_molmerPRL2001}), the physically-allowed region can be obtained by considering all quantum ground states of Hamiltonians of the form $\lambda_1 \hat{\bf J}^2 + \lambda_2 \hat{\bf N}_0^2$ for all values of $\lambda_1, \lambda_2$, and forming the convex envelope of the corresponding data points. The excluded (non-feasible) region is indicated in light grey in Fig.~\ref{fig_spin_1}(a). Regarding the ground state of $\hat{H}_{j=1}$, three limiting cases are especially illustrative. For $q \to +\infty$, the ground state is the polar state $|0,N,0\rangle=\otimes_{i=1}^N|0_z\rangle_i$, which is separable and therefore does not violate any EW. For $q=0$, the ground state is a spin singlet, and violates both our witness Eq.~\eqref{EW_singlet_spin-1} and the SSI $\mathrm{Var}(\hat{\mathbf{J}}) \ge Nj$. In-between, we notice that our EW detects entanglement in the ground state of $\hat H_{j=1}$ for a larger range of values of $q$ than SSIs. Finally, for $q \to -\infty$, the ground state is a twin Fock state $|N/2,0,N/2\rangle$. This state does not violate any EW if only collective spin observables are measured (notice, though, that it does violate a generalized SSI for $j=1/2$ particles, when considering only the two-dimensional subspace spanned by Zeeman sublevels $s=\pm 1$). Yet, the twin-Fock states robustly violates the EW of Eq.~\eqref{EW_singlet_spin-1} (with $\mathrm{Var}(\hat{\mathbf{J}})=N$ and $\langle \hat{\mathbf{N}}_0^2\rangle=N(3N+2)/4$, so that the l.h.s is $\sim N/4$, while the r.h.s is $\sim N/2$). The polar, singlet and twin-Fock states are indicated on Fig.~\ref{fig_spin_1}(a). \\

\paragraph{Ferromagnetic interactions.}
For ferromagnetic interactions ($c<0$) and $q/|c| \lesssim 1.2$ ($N=100$), we find violation of (see Appendix \ref{app_EW}):
\begin{align}
(N-1){\rm Var}(\hat J_z) - N \langle \hat{Q}_z\rangle-\langle
    \hat{J}_x^2 + \hat{J}_y^2 \rangle \rangle \nonumber \\
 +\frac{3N(N+1)}{2}\geq \langle \hat{\mathbf{N}}_0^2\rangle     \label{EW_z_spin-1} 
\ .
\end{align}
Maximal violation is found close to $q=0$ where the state is a Dicke state, such that $\hat{\bf J}^2|{\rm Dicke}\rangle=Nj(Nj+1)|{\rm Dicke}\rangle$ (maximal total spin), and $\hat J_z|{\rm Dicke}\rangle=0$. Eq.~\eqref{EW_z_spin-1} can be compared to another SSI of ref.~\cite{vitaglianoetal2014} violated by the Dicke state, namely $(N-1){\rm Var}(\hat{J}_z)-N \langle \hat Q_z \rangle-\langle \hat J_x^2 + \hat J_y^2\rangle + Nj(Nj+1)\geq 0$. In Fig.~\ref{fig_spin_1}(b), we compare their power in detecting entanglement in the ground state of $\hat H_{j=1}$. In particular, for $q\lesssim |c|$, the ground state of $\hat H_{j=1}$ is very close to the twin-Fock state $|N/2,0,N/2\rangle$, which robustly violates Eq.~\eqref{EW_z_spin-1}: the l.h.s is equal to $N(N+1)/2\sim N^2/2$, while on the r.h.s $\langle \hat{\mathbf{N}}_0^2\rangle = N(3N+2)/4 \sim 4N^2/4$. To conclude on the spin-1 BEC, we find that collective population measurements in orthogonal Zeeman sublevels can reveal entanglement beyond collective spin observables, and our method can be used to infer novel EWs in an optimal and data-agnostic way from these measurements. These new EWs are generically more powerful than SSIs (which can be recovered as a special case in our approach, see Appendix \ref{app_SSI}). In particular, we exhibit two EWs [Eqs.~\eqref{EW_singlet_spin-1} and \eqref{EW_z_spin-1}] robustly violated by the twin-Fock state $|N/2,0,N/2\rangle$, stabilized as the ground state of $\hat H_{j=1}$ in the limit $q \ll |c|$; yet no generalized SSI for spin-1 ensembles is violated by the twin-Fock state.

\begin{figure}[ht]
	\includegraphics[width=\linewidth]{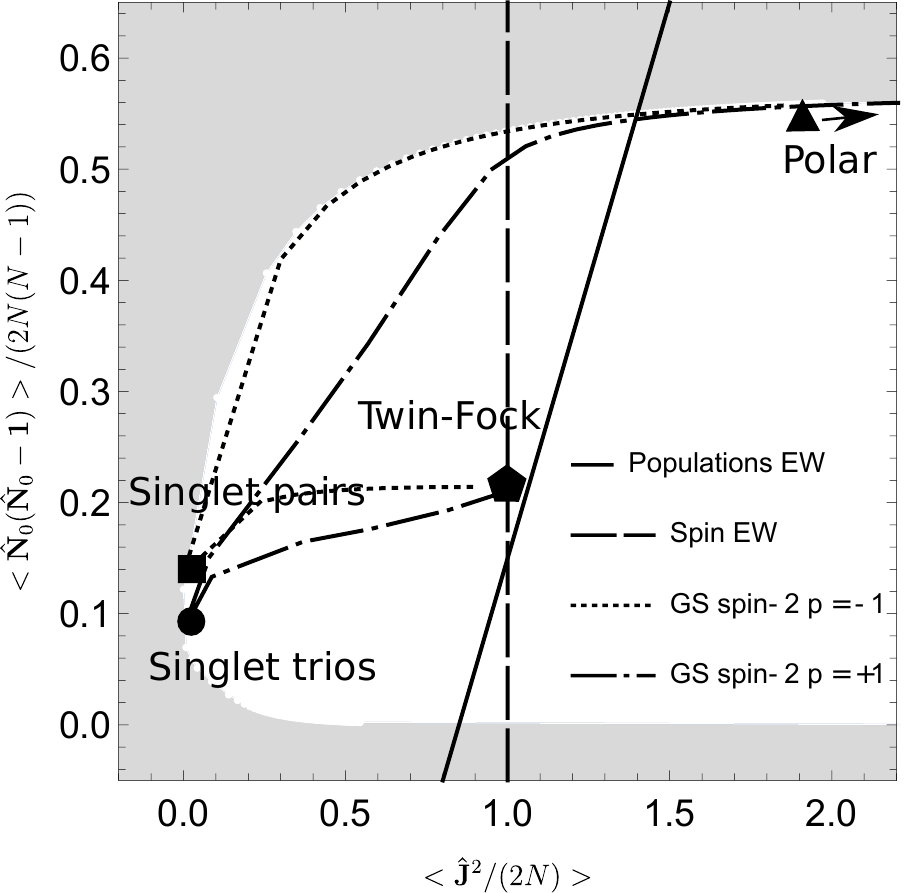}
	\caption{Comparison of the population-based entanglement witness of Eq.~\eqref{EW_singlet_spin-j} (solid line) with the related SSI ${\rm Var}(\hat{\bf J}) \ge Nj$ (dashed line) for spin-2 ensembles ($N=24$). The light grey area are data points which cannot be achieved (see text and Appendix \ref{app_feasible}). Data in the ground state of Eq.~\eqref{H_spin-2} are computed for $c=1$, and $p=1$ (dotted line) or $p=-1$ (dashed-dotted line). Four representative states are indicated: the polar state $|0,0,N,0,0\rangle$ (with $\langle \mathbf{J}^2/(2N)\rangle=3$), the twin-Fock state $|N/2,0,0,0,N/2\rangle$ and two forms of spin singlets (${\rm Var}({\bf J})=0$), obtained either as a pair condensate or a triplet condensate (see text).}
	\label{fig_spin_2}
\end{figure}

\section{Applications to a spin-2 BEC}
The Hamiltonian of a BEC of $j=2$ atoms interacting via pairwise isotropic interactions reads (see Appendix \ref{app_spinor_model}):
\begin{equation}
    \hat{H}_{j=2}=\frac{c}{N}\hat{\mathbf{J}}^2 + \frac{p}{N} \hat{\Theta}_2^\dagger\hat{\Theta}_2 + q\hat{Q}_z\ .
    \label{H_spin-2}
\end{equation}
In contrast to the spin-1 case [Eq.~\eqref{H_spin-1}], the Hamiltonian contains the additional pairing term $\hat{\Theta}_2^\dagger\hat{\Theta}_2$ where $\hat{\Theta}_{2}=\sum_{s=-j}^j(-1)^s\hat{a}_{s}\hat{a}_{-s}$ annihilates time-reversal pairs $\pm s$, with $\hat{a}_s$ annihilating an atom in the spin state $|s_z\rangle$. Notice that $[\hat{\mathbf{J}}^2, \hat{\Theta}_2^\dagger\hat{\Theta}_2] = 0$. We explored the ground state of $\hat{H}_{j=2}$ to find novel violated EW. Using as local operators $\hat S_a(i)$ (the spin) and $\hat n_{a,0}(i)$ (the projector onto the $s=0$ spin state) for $a \in \{x,y,z\}$, the bound in Eq.~\eqref{eq_betaBound} is numerically found as $\beta_2 \approx 4.3$. For this choice of local operators, we found that generalizations of Eqs.~\eqref{EW_singlet_spin-1} and \eqref{EW_z_spin-1} are violated. These EW can be further generalized to arbitrary integer spins, and read (see Appendix \ref{app_EW}):
\begin{equation}
    \mathrm{Var}(\hat{\mathbf{J}})-\frac{1}{N-1}\langle \hat{\mathbf{N}}_0(\hat{\mathbf{N}}_0-\mathbf{1}) \rangle \ge N[j(j+1) - \beta_j]\ ,
      \label{EW_singlet_spin-j} 
\end{equation}
and:
\begin{align}
   (N-1) {\rm Var}(\hat{J}_z)- N \langle \hat Q_z\rangle-\langle \hat J_x^2 + \hat J_y^2 \rangle  \nonumber \\
   +N[(N-1)\beta_j+j(j+1)]\geq \langle \hat{\mathbf{N}}_0(\hat{\mathbf{N}}_0-\mathbf{1}) \rangle ~,
       \label{EW_z_spin-j} 
\end{align}
where $\mathbf{1}=(1,1,1)$ and $\beta_j$ is the corresponding bound in Eq.~\eqref{eq_betaBound}. As for the spin-1 case, one can compare the entanglement detection power of these witnesses with the related SSIs; this comparison is presented in Fig.~\ref{fig_spin_2} for Eq.~\eqref{EW_singlet_spin-j}, while for Eq.~\eqref{EW_z_spin-j} we notice that the improvement with respect to the SSI is mild, and not systematic. Interestingly, as in the case of spin-1, the twin-Fock state $|N/2,0,0,0,N/2\rangle$ (obtained in the limit $q \to -\infty$) violates Eq.~\eqref{EW_singlet_spin-j}, while it does not violate any generalized SSI for spin-2 collective spin observables. As a final exploration of the entanglement patterns in the ground state of a spin-2 BEC, we have focused on many-body singlets ($\langle \hat {\bf J}^2\rangle=0$), stabilized at $q=0$ for antiferromagnetic interactions ($c>0$). In the spin-1 case, there is a unique such state for a BEC, obtained as a ``pair condensate'' $(\hat{\Theta}^{\dagger}_2)^{N/2} \ket{\mathrm{vac.}}$. This pair condensate is also stabilized in the spin-2 case for $p<0$. Yet, for $p>0$, the many-body singlet is instead a ``triplet condensate'' $(\hat{\Theta}^{\dagger}_3)^{N/3} \ket{\mathrm{vac.}}$, where $\Theta^{\dagger}_{3} = -9[(a^\dagger_{-1})^2 a^\dagger_{2} + (a^\dagger_{1})^2 a^\dagger_{-2}] + \sqrt{6} a^\dagger_0 [-(a^\dagger_0)^2 + 3a^\dagger_1 a^\dagger_{-1} + 6a^\dagger_2 a^\dagger_{-2}]$ is the singlet-trio creation operator. Both states are entangled and violate Eq.~\eqref{EW_singlet_spin-j}. However, considering as local measurements $\{ \hat{n}_{a,+2}(i), \hat{n}_{a,+1}(i), \hat{n}_{a,-1}(i), \hat{n}_{a,-2}(i) \}_{a\in\{x,y,z\}}$, we found the following EW violated only by the pair condensate and not by the triplet condensate (see Appendix \ref{app_EW_1}):
\begin{align}
\label{EW_nem}
\langle \hat{\boldsymbol{\sigma}}_2^2+\hat{\boldsymbol{\sigma}}_1^2\rangle-\frac{N-1}{N}(\langle
\hat{\boldsymbol{\sigma}}_2\rangle^2+\langle\hat{\boldsymbol{\sigma}}_1\rangle^2)\nonumber\\
\nonumber
-\frac{2}{N}\langle\hat{\mathbf{N}}_{+2}^2+\hat{\mathbf{N}}_{+1}^2+\hat{\mathbf{N}}_{-1}^2+\hat{\mathbf{N}}_{-2}^2 \rangle\\ +(1-\frac{2}{N})\langle\hat{\mathbf{N}}_{0}\cdot\mathbf{1} \rangle +3\geq 0\ ,
\end{align} 
where we denoted by $\hat{\sigma}_{a,s} = \hat{N}_{a,s} - \hat{N}_{a,-s}$ the population imbalance between opposed levels along direction $a$. Notice that $\{\hat{\sigma}_{x,s},\hat{\sigma}_{y,s},\hat{\sigma}_{z,s}\}$ generate an $su(2)$ subalgebra of $su(5)$. This result illustrates the difference in entanglement properties of two different spin singlets, which by definition cannot be distinguished if only collective spin observables are measured. \\

\section{Conclusions} We have presented a new method to infer violated entanglement witnesses from the collective measurement of Zeeman-sublevel populations in spinor gases of arbitrary spin-$j$ atoms. Our method recovers all known generalized spin squeezing inequalities \cite{vitaglianoetal2014} as a special case, in a data-agnostic way. But when considering appropriate correlations between Zeeman-sublevel populations, one can also construct novel entanglement criteria which have no analog if only the collective spin is measured. We have illustrated our method on the ground states of spin-1 and spin-2 BECs, interacting with contact two-body isotropic interactions in an external magnetic field. In particular, for both $j=1$ and $j=2$ we showed that the twin Fock state robustly violates an entanglement witness involving all populations in three orthogonal directions, while entanglement cannot be detected if only the collective spin is measured. For $j=2$, we presented an entanglement witness distinguishing between two forms of many-body singlets, namely a pair condensate and a triplet condensate. Overall, our new approach appears especially suited to probe entanglement among many particles in high-spin spinor gases, for which entanglement detection has remained a considerable challenge. We would like to emphasized that, although we have focused on collective measurements which are invariant under the permutations of the atoms, our approach is not limited to this situation. It could also be used to probe entanglement in spatially-structured systems by using different local observables $\hat o_k(i)$ for each atom; a straightforward example is to introduce local phases $e^{i\phi_k(i)}$, leaving the bound $\beta$ invariant in Eq.~\eqref{eq_betaBound}. As pointed out recently \cite{frerotetal2022}, this offers considerable flexibility to probe entanglement through components of the structure factors. Finally, a limitation of our current implementation is that there is no optimal and systematic way of incorporating all available first- and second-order moments of population measurements into our main inequality \eqref{eq_central_ineq}. Indeed, while adding more measurement operators \textit{a priori} improves the entanglement detection power, this leads to an increase of the $\beta$ bound in Eq.~\eqref{eq_betaBound}, and can make Eq.~\eqref{eq_central_ineq} harder to violate. The optimization of the choice of measurement operators is left open to future studies.

\paragraph{Code availability.} The code used to diagonalize the spin-1 and spin-2 Hamiltonians, to compute the collective populations in the ground state, and to infer the optimal entanglement witnesses presented in this paper, is available at \url{https://github.com/GuillemMRR/EW_Zeeman}. See Appendix \ref{app_implementation} for a concrete example.

\acknowledgments{We are grateful to Fabrice Gerbier for useful discussions and comments. GM and ML acknowledges support from: ERC AdG NOQIA; Agencia Estatal de Investigación (R$\And$D project CEX2019-000910-S, funded by MCIN/ AEI/10.13039/501100011033, Plan National FIDEUA PID2019-106901GB-I00, FPI, QUANTERA MAQS PCI2019-111828-2, Proyectos de I+D+I “Retos Colaboración” QUSPIN RTC2019-007196-7); Fundació Cellex; Fundació Mir-Puig; Generalitat de Catalunya through the European Social Fund FEDER and CERCA program (AGAUR Grant No. 2017 SGR 134, QuantumCAT U16-011424, co-funded by ERDF Operational Program of Catalonia 2014-2020); EU Horizon 2020 FET-OPEN OPTOlogic (Grant No 899794); National Science Centre, Poland (Symfonia Grant No. 2016/20/W/ST4/00314); European Union’s Horizon 2020 research and innovation programme under the Marie-Skłodowska-Curie grant agreement No 101029393 (STREDCH) and No 847648 (“La Caixa” Junior Leaders fellowships ID100010434: LCF/BQ/PI19/11690013, LCF/BQ/PI20/11760031, LCF/BQ/PR20/11770012, LCF/BQ/PR21/11840013). IF acknowledges support from the Agence Nationale de la Recherche (Qu-DICE project ANR-PRC-CES47), the John Templeton Foundation (Grant No. 61835), European Union’s Horizon 2020 research and innovation programme under the Marie-Skłodowska-Curie grant agreement No 101031549 (QuoMoDys).}

\bibliographystyle{plainnat}
\bibliography{biblio}

\onecolumn\newpage
\appendix
 \section{Derivation of the main entanglement witness inequality}
 \label{app_main_result}
 In this section, we derive our central result, Eq.~\eqref{eq_central_ineq}. In order to detect entanglement from the available data [Eqs.~\eqref{eq_Qdata_1body} and \eqref{eq_Qdata_2body}], our starting point to is to assume that the latter are explained by a separable state, namely a quantum state of the form:
 \begin{equation}
 	\hat \rho_{\rm sep} = \sum_\lambda p_\lambda \otimes_{i=1}^N \hat \rho_\lambda(i) ~,
 	\label{eq_sep_state}
 \end{equation}
 and to arrive at a contradiction. We introduce the notations:
 \begin{align}
 	& m_k(\lambda, i) := {\rm Tr}[\hat \rho_\lambda(i) \hat o_k(i)] \\
 	& m_k(\lambda) := \frac{1}{N} \sum_{i=1}^N m_k(\lambda, i) ~.
 \end{align}
 The key insight of our approach is that, if the state is separable, we may interpret the quantities $m_k(\lambda,i)$ as local classical variables, whose collective fluctuations must reproduce the observed fluctuations of collective observables. The constraints that such local classical variables must obey give rise to our central inequality [Eq.~\eqref{eq_central_ineq}]. Most importantly, the constraints of Eq.~\eqref{eq_betaBound} can be expressed as:
 \begin{equation}
     \frac{1}{N} \sum_{i=1}^N \sum_k [m_k(\lambda, i)]^2 \le \beta
     \label{eq_betaBound_SM}
 \end{equation}
 for all $\lambda$.
 With the above notations, the one-body quantum data [Eq.~\eqref{eq_Qdata_1body}] read:
 \begin{equation}
     m_k = \frac1N \sum_{i=1}^N \langle \hat o_k(i) \rangle = \sum_\lambda p_\lambda m_k(\lambda)
 \end{equation}
 We then introduce the vector notation ${\bf m} := (m_k)$, and the outer product notation $[{\bf u} \otimes {\bf v}]_{ab} = u_a v_b$. With these notations, the two-body quantum data [Eq.~\eqref{eq_Qdata_2body}] read:
 \begin{equation}
 	C = \frac{1}{N(N-1)} \sum_{i \neq j} \sum_\lambda p_\lambda ~ {\bf m}(\lambda, i)^* \otimes {\bf m}(\lambda, j) ~.
 \end{equation}
We now introduce the fluctuations of the $m_k(\lambda,i)$ variables:
\begin{align}
	& A := \sum_\lambda p_\lambda [{\bf m}(\lambda) - {\bf m}]^* \otimes [{\bf m}(\lambda) - {\bf m}] \succeq 0 \\
	& a := \sum_\lambda p_\lambda \frac{1}{N} \sum_i[{\bf m}(\lambda,i) - {\bf m}(\lambda)]^* \otimes [{\bf m}(\lambda,i) - {\bf m}(\lambda)] \succeq 0
\end{align}
The notation $M \succeq 0$ means the the matrix $M$ is positive semidefinite (PSD), that is, for any vector ${\bf u}$, ${\bf u}^\dagger M {\bf u} \ge 0$. This PSD property is straightforward to verify, since $A$ and $a$ are both of the form $\sum_x w_x {\bf v} {\bf v}^\dagger$ with $w_x \ge 0$ and ${\bf v}$ a one-dimensional vector.

The matrix $C$ is then:
\begin{align}
	 C &= \frac{N}{N-1}[{\bf m}^* \otimes {\bf m} + A] - \frac{1}{N(N-1)} \sum_\lambda p_\lambda \sum_i {\bf m}(\lambda,i)^* \otimes {\bf m}(\lambda, i) \label{def_C_0}\\
	&= {\bf m}^* \otimes {\bf m} + A - \frac{1}{N-1} a ~. \label{def_C}
\end{align}	
From Eq.~\eqref{def_C_0} and Eq.~\eqref{eq_betaBound_SM}], we have:
\begin{equation}
	{\rm Tr}(C) \ge \frac{N}{N-1}{\rm Tr}[{\bf m}^* \otimes {\bf m} + A] - \frac{\beta}{N-1} ~.
	\label{eq_ineq1_C}
\end{equation}
We then introduce a projector $P$, and write:
\begin{align}
	{\rm Tr}[{\bf m}^* \otimes {\bf m} + A] &= {\rm Tr}[P({\bf m}^* \otimes {\bf m} + A)] + {\rm Tr}[(1-P)({\bf m}^* \otimes {\bf m} + A)] \\
	&= {\rm Tr}[P({\bf m}^* \otimes {\bf m} + A)] + {\rm Tr}[(1-P)(C + a/(N-1))] 
\end{align}
where we used Eq.~\eqref{def_C}.
From the positivity of $A$ and $P$, we have ${\rm Tr}(PA) \ge 0$. Similarly, from the positivity of $a$ and $1-P$ (which is also a projector), we have ${\rm Tr}[(1-P)a] \ge 0$. Hence:
\begin{equation}
	{\rm Tr}[{\bf m}^* \otimes {\bf m} + A] \ge {\rm Tr}[P({\bf m}^* \otimes {\bf m})] + {\rm Tr}[(1-P)C]
\end{equation}
Introducing this inequality into Eq.~\eqref{eq_ineq1_C}, and reorganizing the terms, this leads to our central result:
\begin{equation}
	 L(P): = N ~{\rm Tr}[ P( C - {\bf m}^* \otimes {\bf m})]  \ge {\rm Tr}(C)  - \beta ~,
	 \label{eq_central_ineq_suppmat}
\end{equation}
which is valid for any projector $P$ whenever the state is separable [Eq.~\eqref{eq_sep_state}]. This central inequality is a generalization of the results contained of Vitagliano \textit{et al.} \cite{vitaglianoetal2014}. In order to detect entanglement at best, one has to find the optimal projector $P$ such that inequality \eqref{eq_central_ineq_suppmat} is violated -- which would invalidate our starting assumption, namely that the state is separable. This optimization is straightforward: in order to minimize $L$ over the projector $P$, one has to choose $P$ as the projector onto the subspace of negative eigenvalues of $C - {\bf m}^* \otimes {\bf m}$.

\section{Recovering the generalized spin-squeezing inequalities of ref.~\cite{vitaglianoetal2014}.}
\label{app_SSI}
In ref.~\cite{vitaglianoetal2014}, a family of SSI was derived that can be obtained as a special case of our approach. These inequalities are valid for all fully-separable states of $N$ spin-$j$ atoms, and involve first and (modified) second moments of collective spins in three orthogonal directions: $\hat J_x, \hat J_y, \hat J_z$. We write here these inequalities (Eq.~(9) of ref.~\cite{vitaglianoetal2014}) for completeness:
\begin{eqnarray}
    \langle \hat J_x^2 + \hat J_y^2 + \hat J_z^2 \rangle & \le & Nj(Nj+1) \label{eq_vitagliano_9a}\\
    {\rm Var}(\hat J_x) + {\rm Var}(\hat J_y) + {\rm Var}(\hat J_z) &\ge& Nj \label{eq_vitagliano_9b}\\
    \langle \hat J_l^2 \rangle - \sum_{i=1}^N \langle [\hat S_l(i)]^2 \rangle + \langle\hat J_m^2 \rangle - \sum_{i=1}^N \langle [\hat S_m(i)]^2 \rangle &\le& (N-1) \left[{\rm Var}(\hat J_k) - \sum_{i=1}^N \langle [\hat S_k(i)]^2 \rangle \right] \nonumber\\&&+  N(N-1)j^2 \label{eq_vitagliano_9c}\\
    \langle \hat J_m^2 \rangle - \sum_{i=1}^N \langle [\hat S_m(i)]^2 \rangle -  N(N-1)j^2 &\le& (N-1) \left[{\rm Var}(\hat J_k) - \sum_{i=1}^N \langle [\hat S_k(i)]^2 \rangle \right.\nonumber\\&&\left.+ {\rm Var}(\hat J_l) - \sum_{i=1}^N \langle [\hat S_l(i)]^2 \rangle \right] \label{eq_vitagliano_9d}
\end{eqnarray}
The labels $k,l,m$ in the last two inequalities may take all permutations of $x,y,z$. As we shall show, these inequalities are obtained from our main inequality Eq.~\eqref{eq_central_ineq} (or equivalently Eq.~\eqref{eq_central_ineq_suppmat} in Appendix \ref{app_main_result}), using as local observables $\hat o_k(i)$ the three spin observables $\hat S_x(i), \hat S_y(i), \hat S_z(i)$. We then have:
\begin{eqnarray}
    m_k &=& \langle \hat J_k \rangle / N \\
    C_{kl} &=& \frac{1}{N(N-1)} \left[\langle \hat J_k \hat J_l \rangle - \sum_i \langle \hat S_k(i) \hat S_l(i) \rangle\right] ~.
\end{eqnarray}
Furthermore, the $\beta$ bound is $\beta=j^2$, since for any state of a spin-$j$ atom, one has $\langle S_x(i) \rangle^2 + \langle S_y(i) \rangle^2 + \langle S_z(i) \rangle^2 \le j^2$, with equality if the atomic spin is fully polarized along some direction. Hence we have:
\begin{equation}
     {\rm Tr}(C)  - \beta = \frac{1}{N(N-1)} (\langle J_x^2 + J_y^2 + J_z^2 \rangle) - \frac{1}{N-1} j(j+1) -j^2
\end{equation}
where we used that $\langle [S_x(i)]^2 +  [S_y(i)]^2 +  [S_z(i)]^2 \rangle = j(j+1)$. As we will show explicitly, the 8 SSI [Eqs.~\eqref{eq_vitagliano_9a} to \eqref{eq_vitagliano_9d}] are obtained by choosing the projector $P$ in  Eq.~\eqref{eq_central_ineq_suppmat} diagonal, leading to inequalities of the form:
\begin{equation}
    N\sum_{k \in E} [C_{kk} - m_k^2] \ge {\rm Tr}(C) - j^2
\end{equation}
with $E$ any subset of $\{x,y,z\}$. This last inequality can be written more explicitly as:
\begin{equation}
    \sum_{k \in E} \left[N \langle \hat J_k^2 \rangle -(N-1)\langle \hat J_k \rangle^2 - N \sum_i \langle [S_k(i)]^2\rangle\right] \ge \langle J_x^2 + J_y^2 + J_z^2 \rangle - Nj(Nj+1) ~.\label{eq_central_ineq_vitagliano}
\end{equation}
Choosing $E=\emptyset$ directly gives Eq.~\eqref{eq_vitagliano_9a}.\\
Choosing $E=\{x,y,z\}$ gives:
\begin{equation}
    (N-1)\left[{\rm Var}(\hat J_x) +  {\rm Var}(\hat J_y) +  {\rm Var}(\hat J_z)\right] \ge Nj(j+1) - Nj(Nj+1) ~,
\end{equation}
namely Eq.~\eqref{eq_vitagliano_9b}.\\
To obtain Eqs.~\eqref{eq_vitagliano_9c} and \eqref{eq_vitagliano_9d}, we rewrite Eq.~\eqref{eq_central_ineq_vitagliano} as:
\begin{eqnarray}
    \sum_{k \in E} \left[N \langle \hat J_k^2 \rangle -(N-1)\langle \hat J_k \rangle^2 - N \sum_i \langle [S_k(i)]^2\rangle\right] &\ge & \langle J_x^2 + J_y^2 + J_z^2 \rangle - \sum_i\left\langle [\hat S_x(i)]^2 + [\hat S_y(i)]^2 + [\hat S_z(i)]^2 \right\rangle \nonumber\\&&- N(N-1)j^2 ~.
\end{eqnarray}
Taking then e.g. $E=\{x\}$, we find:
\begin{equation}
    (N-1)\left[{\rm var}(\hat J_x) - \sum_i\langle [\hat S_x(i)]^2 \rangle \right] \ge \langle J_y^2 + J_z^2 \rangle - \sum_i\left\langle [\hat S_y(i)]^2 + [\hat S_z(i)]^2 \right\rangle - N(N-1)j^2 ~,
\end{equation}
namely Eq.~\eqref{eq_vitagliano_9c} for $(k,l,m)=(x,y,z)$, and similarly for $E=\{y\}$ or $E=\{z\}$.\\
Finally, Eq.~\eqref{eq_vitagliano_9d} is obtained by choosing $E=\{x,y\}$, $E=\{x,z\}$ or $E=\{y,z\}$.\\
This concludes the proof that the generalized SSI of ref.~\cite{vitaglianoetal2014} are incorporated in the framework presented in this paper when choosing the spin components as local observables.

\section{The spinor Bose gas model}
\label{app_spinor_model}
In this section, we present some background information, based on Ref.~\cite{KAWAGUCHI2012253}, on the BEC Hamiltonians used in the main text [Eqs.~\eqref{H_spin-1} and \eqref{H_spin-2}]. We consider an ensemble of spin-$j$ bosonic atoms with contact, two-body, spin-conserving interactions in $\mathbb{R}^3$. In the single-mode approximation, its Hamiltonian reads:
\begin{equation}
    \hat{H}_0(\{g_f\}_{f=0}^j)=\sum_{s_1s_2,s_3s_4=-j}^jg_fC_{s_1s_2,s_3s_4}^{j\otimes j=2f}\hat{a}^\dagger_{s_1}\hat{a}^\dagger_{s_2}\hat{a}_{s_3}\hat{a}_{s_4} \ ,
    \label{H_spinor_isotropic}
\end{equation}
where $g_f$ is the scattering amplitude of the spin-$2f$ channel. The matrix element of the projection of a composite spin-$2f$ onto two spin-$j$ atoms is $C_{s_1s_2,s_3s_4}^{j\otimes j=2f}=\bra{j,s_1}_j\otimes \bra{j,s_2}_j\mathcal{P}^{j\otimes j=2f} \ket{j,s_3}_j\otimes \ket{j,s_4}_j$, with $\mathcal{P}^{j\otimes j=2f}=\sum_{S=-2f}^{2f}\ket{2f,S}_{j\otimes j}\bra{2f,S}_{j\otimes j}$, where $S$ is the spin projection of the composite spin-$2f$. The operators $\hat{a}_{s}(\hat{a}^\dagger_{s})$ annihilate (create) a bosonic mode with spin projection $s\in \{-j,-j+1,..,j  \}$.  The total number of atoms $\hat{N}=\sum_{s=-j}^j\hat{N}_{s}$, where $\hat{N}_{s}=\hat{a}^\dagger_{s}\hat{a}_{s}$, is fixed to $N$. 
 
In the following we will express the Hamiltonian  \eqref{H_spinor_isotropic} in terms of collective operators. This is done via the identification: 
\begin{equation}
    \hat{O}= \sum_{s_L, s_R=-j}^j \bra{s_L}\hat{o}\ket{s_R} \hat{a}^\dagger_{s_L}\hat{a}_{s_R} \ ,
    \label{collective_bose}
\end{equation}

where $\hat{O}=\sum_{i=1}^N{\hat{o}}(i)$ is the collective operator corresponding to the local observable $\hat{o}$ acting on a single-atom spin-$j$ degree of freedom. Note that in  particular the population operator $\hat{N}_s$ is a collective operator with $\bra{s_L}\hat{n}_s\ket{s_R}:=\braket{s_L}{s}\braket{s}{s_R}=\delta_{s_L,s}\delta_{s,s_R}$.

\paragraph{Spin 1.} For spin-1, the spin projection onto the cartesian basis $\{x,y,z\}$ reads: 

\begin{equation}
    \hat{J}_x= \frac{1}{\sqrt{2}} 
    \begin{pmatrix}
    \hat{a}_{+1}& \hat{a}_0 &\hat{a}_{-1} \\
    \end{pmatrix}
    \begin{pmatrix} 
    0 & 1 & 0 \\
    1 & 0 & 1 \\
    0 & 1 & 0 \\
    \end{pmatrix}
        \begin{pmatrix}
    \hat{a}_{+1} \\
    \hat{a}_0 \\
    \hat{a}_{-1}
    \end{pmatrix}
\end{equation}
\begin{equation}
    \hat{J}_y= \frac{1}{\sqrt{2}} 
    \begin{pmatrix}
    \hat{a}_{+1}& \hat{a}_0 &\hat{a}_{-1} \\
    \end{pmatrix}
    \begin{pmatrix} 
    0 & -\mathbb{i} & 0 \\
    \mathbb{i} & 0 & -\mathbb{i} \\
    0 & \mathbb{i} & 0 \\
    \end{pmatrix}
        \begin{pmatrix}
    \hat{a}_{+1} \\
    \hat{a}_0 \\
    \hat{a}_{-1}
    \end{pmatrix}
\end{equation}
\begin{equation}
    \hat{J}_z= 
    \begin{pmatrix}
    \hat{a}_{+1}& \hat{a}_0 &\hat{a}_{-1} \\
    \end{pmatrix}
    \begin{pmatrix} 
    1 & 0 & 0 \\
    0 & 0 & 0 \\
    0 & 0 & -1 
    \end{pmatrix}
        \begin{pmatrix}
    \hat{a}_{+1} \\
    \hat{a}_0 \\
    \hat{a}_{-1}
    \end{pmatrix} \ ,
\end{equation}
where $\mathbb{i}$ is the imaginary unit, $\mathbb{i}^2=-1$. With this expressions, we can compute the total spin, using the bosonic commutation relations and $N_{z,+1}+N_{z,0}+N_{z,-1}=N$ we have: 
\begin{align}
\nonumber
    \hat{\mathbf{J}}^2:&= \hat J_x^2 + \hat J_y^2 + \hat J_z^2= \\ &=\hat{J}_z^2+ \hat{N}+\hat{N}_0+2\hat{N}_0(N-\hat{N}_0)+2((\hat{a}^\dagger_{0})^2\hat{a}_{+1}\hat{a}_{-1}+\mathrm{h.c.})
\end{align}

Alternatively, by identifying the pair annihilation operator $\hat{\Theta}_{2}=\hat{a}_0^2-2\hat{a}_{+1}\hat{a}_{-1}$, we can write $\hat{\textbf{J}}^2=N^2-\hat{\Theta}^\dagger\hat{\Theta}$. As first observation, we notice that unlike the spin-1/2 case, $\hat{\mathbf{J}}^2$ is not diagonal in the occupation basis. Direct computation shows that the Hamiltonian \eqref{H_spinor_isotropic} is equivalent to: 
\begin{equation}
    \hat{H}_0(c_{1})=c_{1}\hat{\mathbf{J}}^2\ ,
    \label{H_spin1_iso}
\end{equation}
where $c_1=(g_1-g_0)/3$. 

\paragraph{Spin 2.}

For spin-2, there is a second relevant interaction channel. In this case, the Hamiltonian reads:   
\begin{equation}
    \hat H_0(c_2,p) = c_2 \hat {\bf J}^2 + p[ \hat{N}_0(\hat{N}_0-1)+4(\hat{N}_{1}\hat{N}_{-1}+\hat{N}_{-2}\hat{N}_{2})+2(\hat{T}_{0,-2}^+\hat{T}_{0,2}^+-\hat{T}_{0,-1}^+\hat{T}_{0,1}^+-2\hat{T}_{1,2}^ +\hat{T}_{-1,-2}^++(+\leftrightarrow-))] \ ,
\end{equation}
where $c_2=(g_4-g_2)/7, p=(7g_0-10g_2+3g_4)/35$ and we introduced the collective operators $\hat{T}_{s,s'}^{+}=\hat{a}^\dagger_s\hat{a}_{s'}+\mathrm{h.c.},\hat{T}_{s,s'}^{-}=\mathbb{i}(\hat{a}^\dagger_s\hat{a}_{s'}-\mathrm{h.c.}) $. Using the pair anhiliation operator for spin-2, $\hat{\Theta}_2=\hat{a}_0^2-2\hat{a}_1\hat{a}_{-1}+2\hat{a}_{+2}\hat{a}_{-2}$, the term coupling $p$ can be interpreted as a pair amplitude $\hat{\Theta}_2^\dagger\hat{\Theta}_2$. This term commutes with $\hat{\mathbf{J}}^2$, although we do not have written it explicitly in a rotationally invariant way using cartesian coordinates.   

\paragraph{Quadratic Zeeman shift.} In this work we consider the gas is immersed in a uniform magnetic field of intensity $B$ along the $z$ direction. In this case, the total Hamiltonian becomes $\hat H = \hat H_0 / N + q \hat{Q}_z$, where $q\propto B^2$ and $\hat{Q}_z=\sum_{i=1}^N\hat{J}_z^2$. We introduce a scaling to make the Hamiltonian extensive in $N$. We consider an experiment where all atoms are initially prepared in the state $\otimes_{i=1}^N |0_z\rangle$, so that we may remove the linear Zeeman term, proportional to $\hat J_z$ (as $\hat J_z$ is a symmetry of $\hat H$, and acts trivially on the state).

\section{Computation of the quantum data }
\label{Appendix_quantumdata}
In this appendix, we provide technical details on the computation of the quantum data used in the illustration of our method. Specifically, one has to compute one and two body Zeeman sublevel correlations of the form: 
\begin{equation}
    \langle \hat{N}_{a,s} \rangle \ , \langle \hat{N}_{a,s}\hat{N}_{a,s'} \rangle \ ,\mathrm{etc.} \ ,
\end{equation}
in the ground state of relevant Hamiltonians of spin-$j$ ensembles.

\paragraph{Diagonalization of the Hamiltonians.} The Hamiltonians considered in this paper are polynomials of collective operators $\hat{O}_k=\sum_{i=1}^N\hat{o}_k(i)$ with the spin projection $\hat J_z$ a symmetry. Since operators acting in different atoms commute, we notice that the collective operators $\{\hat O_k\}$ obey the same algebra as their their local counterparts, $\{\hat{o}_k(i)\}$. Accordingly, the problem reduces to compute the matrices of the irreducible representations (irreps) of the algebra, describing the collective operators. The local operators can be described by $d\times d$ Hermitian matrices, where $d=2j+1$. Let $\{\hat{\lambda}_\alpha\}_{\alpha=1}^{d^2-1}$ be a basis of $d\times d$ Hermitian traceless matrices fulfilling the orthogonality condition $\mathrm{Tr}[\hat{\lambda}_\alpha\hat{\lambda}_\beta]=\delta_{\alpha\beta}$, for instance the so-called generalized Gell-Mann matrices. The commutation relations:

\begin{equation}
\hat{\lambda}_\alpha\hat{\lambda}_\beta-\hat{\lambda}_\beta\hat{\lambda}_\alpha=\mathbb{i}\sum_{\gamma=1}^{d^2-1}f_{\alpha\beta\gamma}\hat{\lambda}_\gamma \ ,
\end{equation}

are characterized by the structure constants $f_{\alpha\beta\gamma}$. The structure constants are the same for the generators of the $SU(d)$ group with the Lie bracket $[\hat{\lambda}_\alpha, \hat{\lambda}_\beta]:=\hat{\lambda}_\alpha\hat{\lambda}_\beta-\hat{\lambda}_\beta\hat{\lambda}_\alpha$. The representations of $SU(d)$ are labelled by partitions of $N$ in $d$ parts as the permutation group of $N$ $d$-level atoms. In the totally symmetric sector with respect to the exchange of atoms, the matrix representation of the generators is computed $\{\hat{\Lambda}_\alpha =I[\hat{\lambda}_\alpha]\}_{\alpha=1}^{d^2-1}$. The operator $\hat{\Lambda}_\alpha$ is the collective counterpart of $\hat{\lambda}_\alpha$, that is $\hat{\Lambda}_\alpha=\sum_{i=1}^N\hat{\lambda}_\alpha(i)$. Consequently, for any other operator $\hat{o}$, the map $I$ distributes linearly 

\begin{equation}
    I[\hat{o}]= N\mathrm{Tr}[\hat{o}]\frac{\mathbb{I}}{D}+\sum_{\alpha=1}^{d^2-1}\mathrm{Tr}[\hat{\lambda}_\alpha\hat{o}]\hat{\Lambda}_\alpha:=\hat{O} \ ,
    \label{collective_irrep}
\end{equation}

where $D$ is the dimension of the representation, and the identity element $\mathbb{I}$ is added in order to account for the trace of $\hat{o}$. The multiplicity $N$ is introduced to consistently transform the identity, $ I[\mathbb{1}]=\sum_{i=1}^N\mathbb{1} = N$. The expression \eqref{collective_irrep} is an alternative of the second-quantization form \eqref{collective_bose} providing the advantage that higher moments are computed just by multiplying matrices, as specified in the following.

\paragraph{Gathering the correlations.} The Hamiltonian is diagonalized using the previous matrix representation, allowing one to find the ground state $\ket{\rm GS}$. Then, the computation of moments reduces to matrix multiplication matrices and computing expectation values in the ground state: $\langle \hat{O}_a \rangle = \bra{\rm GS} I[\hat{o}_a] \ket{\rm GS}$, $\langle \hat{O}_a \hat{O}_b \rangle = \bra{\rm GS} I[\hat{o}_a] I[\hat{o}_b] \ket{\rm GS}$, etc, for any collective operators $\hat{O}_a$. In this work in particular we probe the GS with magnetic-sublevel populations along a direction $a$, $\{\hat{N}_{a,s}\}_{s=-j}^j$. These are generated from the single-atom observable $\hat{n}_{a,s}=\prod_{s\neq s'=-j}^j(s-\hat{S}_a)/(s-s')$, where $\hat{S}_a$ is the atomic spin projection along direction $a \in \{x,y,z\}$, i.e $\hat S_a:=\mathbf{a}\cdot\hat{\mathbf{S}}$.

\section{Computation of the physically feasible region of a set of average values}
\label{app_feasible}
In this section, we explain how to compute the feasible region showed in the figures of the main text. Formally, given two observables $\hat X$ and $\hat Y$, acting on the Hilbert space for $N$ qudits, our goal is find the possible combinations of ``feasible'' mean values $(x=\langle \hat X\rangle, y=\langle \hat Y \rangle)$, that is, combinations of mean values which can be obtained in some (generally mixed) quantum state. Equivalently, one may look for the extremal values of $\langle \hat Y \rangle$, while keeping fixed the mean value $x=\langle \hat X\rangle$. By making statistical mixtures of the corresponding quantum states, the full convex hull of these extremal points is also feasible. This problem can be solved introducing a Lagrange multiplier $\lambda_x$, and finding the ground state of Hamiltonians of the form:
\begin{equation}
    \hat L(\lambda_x) = \pm \hat Y - \lambda_x(\hat X - x \mathbb{1}) ~.
\end{equation}
Adjusting the multiplier $\lambda_x$ ensures that the constraint $\langle \hat X \rangle = x$ is met in the ground state. More generally, the extremal points $(x,y)$ are found as the expectation values of $\hat X, \hat Y$ in the ground state of:
\begin{equation}
    \hat L(\lambda_x, \lambda_y) = \lambda_x \hat X + \lambda_y \hat Y ~,
\end{equation}
varying the parameters $\lambda_x, \lambda_y$ between $-\infty$ and $+\infty$. By forming statistical mixtures of these ground states, the full convex hull of the extremal points are also feasible, which leads to the construction showed in the figures of the main text.

In order to find the ground state of $\hat{L}$, the same technique outlined above for the Bose gas Hamiltonian can be applied since in the cases of interest here, $\hat{L}$ is a polynomial of collective operators.

\section{Derivation of the entanglement witness inequalities \eqref{EW_singlet_spin-1}, \eqref{EW_z_spin-1} and their generalization to arbitrary integer spins Eqs.~\eqref{EW_singlet_spin-j}, \eqref{EW_z_spin-j}}
\label{app_EW}
These inequalities are found by choosing, as local measurements, the spin observables $\hat S_a(i) = \sum_{s=-j}^j s \hat n_{a,s}(i)$, as well as the projectors $\hat n_{a,0}(s)$ onto the $s=0$ magnetic sublevels, along the three orthogonal directions $a\in \{x,y,z\}$. For completeness, we rewrite here the EW inequalities \eqref{EW_singlet_spin-j} and \eqref{EW_z_spin-j} of the main text:
\begin{equation}
    \mathrm{Var}(\hat{\mathbf{J}})-\frac{1}{N-1}\langle \hat{\mathbf{N}}_0(\hat{\mathbf{N}}_0-\mathbf{1}) \rangle \ge N[j(j+1) - \beta_j]\ ,
      \label{EW_singlet_spin-j_SM} 
\end{equation}
and:
\begin{equation}
   (N-1) {\rm Var}(\hat{J}_z)- N \langle \hat Q_z\rangle-\langle \hat{J}_x^2 + \hat{J}_y^2 \rangle
   +N[(N-1)\beta_j+j(j+1)]\geq \langle \hat{\mathbf{N}}_0(\hat{\mathbf{N}}_0-\mathbf{1}) \rangle ~,
       \label{EW_z_spin-j_SM} 
\end{equation}
where $\mathbf{1}=(1,1,1)$ and $\beta_j$ is the corresponding bound in Eq.~\eqref{eq_betaBound}. We evaluated this bound numerically, and the corresponding values are:
\begin{table}[h]
  \centering
	\begin{tabular}{| c | c | c | c | c | c |}
	\hline
	  spin $j$ & $1$ & $2$ & $3$ &$4$  & $5$\\
	  \hline
	 $\beta$ $\{\mathbf{S}, \mathbf{n}_0 \}$ & 1.5 & 4.3& 9.2& 16.15 & 25.22\\
	 \hline
	 \end{tabular}
	\caption{Numerical separable bound $\beta$ for the basis of local measurements $\{\{\hat{S}_a(i),\hat{n}_{a,0}(i) \} _{a=\{x,y,z\}}\}_{i=\{1,2,..,N\}}$ for different spin-$j$.   }
	\label{table_bound}
\end{table}\\
In the specific case of $j=1$, the EW inequalities can be simplified using the identities: $\sum_{a\in\{x,y,z\}}(\hat{N}_{a,+1}+\hat{N}_{a,-1})=3N-\sum_{a\in\{x,y,z \}}\hat{N}_{a,0}=\sum_{i=1}^N [\hat{\mathbf{S}}(i)]^2=j(j+1)N=2N$. Together with the bound $\beta_{j=1}=3/2$, we obtain Eqs.~\eqref{EW_singlet_spin-1}, \eqref{EW_z_spin-1} from respectively Eqs.~\eqref{EW_singlet_spin-j_SM} and \eqref{EW_z_spin-j_SM}.

\paragraph{Derivation of ${\rm Tr}[C]$.} To derive these EWs from our central inequality, Eq.~\eqref{eq_central_ineq}, one has to compute ${\rm Tr}[C]$, which reads:
\begin{align}
    \mathrm{Tr}[C] =& \frac{1}{N(N-1)} \sum_{i \neq j} \sum_{a \in \{x,y,z\}} \left\langle\hat{S}_{a}(i) \hat{S}_{a}(j) + \hat{n}_{a,0}(i) \hat{n}_{a,0}(j)\right\rangle \\
    =& \frac{1}{N(N-1)} \sum_{a \in \{x,y,z\}} \left\langle [\hat{J}_a]^2 + [\hat{N}_{a,0}]^2 -  \sum_{i=1}^N [[\hat{S}_a(i)]^2 + \hat{n}_{a,0}(i) ]
    \right\rangle\\
    =&\frac{1}{N(N-1)}\left(\langle \hat{\mathbf{J}}^2+\hat{\mathbf{N}}_0(\hat{\mathbf{N}}_0-\mathbf{1})\rangle-j(j+1)N\right) \label{eq_traceC}
\end{align}
where $\hat {\bf J} = (\hat J_x, \hat J_y, \hat J_z)$ is the collective spin, $\hat{\bf N}_0=(\hat N_{x,0}, \hat N_{y,0}, \hat N_{z,0})$ and ${\bf 1}=(1,1,1)$. We used that $\sum_a [\hat{S}_a(i)]^2 = j(j+1)$, and that $[\hat{n}_{a,0}(i)]^2=\hat{n}_{a,0}(i)$. 
The two EW are obtained by choosing two different projectors in Eq.~\eqref{eq_central_ineq}. Recall that, since correlations in different orientations are not observed, $P$ has a block-diagonal form $(P_x, P_y, P_z)$.

\paragraph{Derivation of Eq.~\eqref{EW_singlet_spin-j_SM}.} For Eq.~\eqref{EW_singlet_spin-j_SM} (tailored in particular to a spin-$j$ many-body singlet), the optimal projection $P$ is $P_x=P_y=P_z= \begin{pmatrix}
1 & 0 \\
0& 0
\end{pmatrix} := P_{j}$ in the local basis described before, leading to
\begin{align}
    \mathrm{Tr}[NP(C-\mathbf{m}\otimes \mathbf{m})]=&\frac{1}{N-1}\left(\langle\hat{\mathbf{J}}^2 \rangle-j(j+1)N\right)-\frac{1}{N}\langle \hat{\mathbf{J}} \rangle^2 \ .
\end{align}
Inserting this expression, together with Eq.~\eqref{eq_traceC}, into our central inequality \eqref{eq_central_ineq}, we obtain the announced result Eq.~\eqref{EW_singlet_spin-j_SM}.

\paragraph{Derivation of Eq.~\eqref{EW_z_spin-j_SM}.}
For Eq.~\eqref{EW_z_spin-j_SM} (tailored in particular to the Dicke states $\mathbf{J^2}=Nj(Nj+1)$, $J_z=0$), the optimal projector is
$P_x=P_y=0, P_z=P_j$, from which we find:
\begin{align}
    \mathrm{Tr}[NP(C-\mathbf{m}\otimes \mathbf{m})]=&\frac{1}{N-1}\langle(\hat{J}_z)^2-\hat{Q}_z \rangle-\frac{1}{N}\langle \hat{J}_z \rangle^2 \ .
\end{align}
Inserting this expression, together with Eq.~\eqref{eq_traceC}, into our central inequality \eqref{eq_central_ineq}, we obtain the announced result Eq.~\eqref{EW_z_spin-j_SM}.

\section{Derivation of spin-2 EW, inequality \eqref{EW_nem} and generalization to arbitrary integer spins}
\label{app_EW_1}

In this appendix we show how to derive the entanglement witness presented in Eq.~\eqref{EW_nem}, and violated by the pair condensate, but not by the triplet condensate. In the present case, we use as local observables the projectors onto all magnetic sublevels, except the $s=0$ sublevel: $\{\hat{n}_{a,+2}(i),\hat{n}_{a,+1}(i),\hat{n}_{a,-1}(i),\hat{n}_{a,-2}(i)   \}_{a\in\{x,y,z\}}$. This choice leads to the $\mathrm{Tr}(C)$ term:
\begin{align}
\mathrm{Tr}(C)=& \frac{1}{N(N-1)}\sum_{a\in\{x,y,z\}} \sum_{s \neq 0} \langle  [\hat N_{a,s}]^2 -\hat N_{a,s} \rangle \\
& \frac{1}{N(N-1)}\sum_{a\in\{x,y,z\}} \left(
    \sum_{s \neq 0} \langle  [\hat N_{a,s}]^2 \rangle - (N - \langle \hat N_{a,0} \rangle
    \right)\\
=& \frac{1}{N(N-1)}\sum_{a\in\{x,y,z\}} \left(
    \sum_{s \neq 0} \langle  [\hat N_{a,s}]^2 \rangle + \langle \hat N_{a,0} \rangle \right) - \frac{3}{N-1}
\end{align}

The method reveals the optimal projector $P_x=P_y=P_z=\frac{1}{2}\begin{pmatrix}
1 & 0 & 0 &-1 \\
0& 1 & -1 &0 \\
0 & -1 & 1& 0 \\
-1& 0& 0& 1 
\end{pmatrix}$. This yields the terms: 
\begin{equation}
    \mathrm{Tr}[NPC]=\frac{1}{2(N-1)}\left(\sum_{a\in\{x,y,z\}} \left\langle [\hat{N}_{a,+2}-\hat{N}_{a,-2}]^2+[\hat{N}_{a,+1}-\hat{N}_{a,-1}]^2 + \hat{N}_{a,0} \right\rangle-3N\right)
\end{equation}
\begin{equation}
    \mathrm{Tr}[NP(\mathbf{m}\otimes\mathbf{m})]=\frac{1}{2N}\sum_{a\in\{x,y,z\}}\left(\langle \hat{N}_{a,+2}-\hat{N}_{a,-2}\rangle^2+\langle\hat{N}_{a,+1}-\hat{N}_{a,-1}\rangle^2 \right)\ .
\end{equation}
We introduce the notation $\hat{\sigma}_{a,s} = \hat{N}_{a,s} - \hat{N}_{a,-s}$, the population imbalance between opposed levels along direction $a$. Together with the notation ${\bf 1}=(1,1,1)$, we can rewrite these equations as:
\begin{equation}
    2(N-1){\rm Tr}(C) = \frac{2}{N} \sum_{s \neq 0} \langle [\hat{\bf N}_s]^2 \rangle + \frac2N \langle \hat {\bf N}_0 \cdot {\bf 1} \rangle - 6
\end{equation}
\begin{equation}
    2N(N-1)\mathrm{Tr}[PC]= \langle [\hat {\boldsymbol{\sigma}}_1]^2 + [\hat {\boldsymbol{\sigma}}_2]^2 \rangle  + \langle \hat {\bf N}_0 \cdot {\bf 1} \rangle - 3N
\end{equation}
\begin{equation}
    2N(N-1)\mathrm{Tr}[P(\mathbf{m}\otimes\mathbf{m})]=\frac{N-1}{N}[\langle \hat {\boldsymbol{\sigma}}_1 \rangle^2 + \langle \hat {\boldsymbol{\sigma}}_2 \rangle^2] .
\end{equation}
Together with the bound $\beta=3/2$ found numerically, we find:
\begin{align}
    2(N-1)\left[N\mathrm{Tr}[P(C - \mathbf{m}\otimes\mathbf{m})] - {\rm Tr}(C) + \beta\right] = & \langle [\hat {\boldsymbol{\sigma}}_1]^2 + [\hat {\boldsymbol{\sigma}}_2]^2 \rangle \nonumber - \frac{N-1}{N}[\langle \hat {\boldsymbol{\sigma}}_1 \rangle^2 + \langle \hat {\boldsymbol{\sigma}}_2 \rangle^2] \nonumber\\
    &+ \left(1 - \frac2N\right)\langle \hat {\bf N}_0 \cdot {\bf 1} \rangle - \frac{2}{N} \sum_{s \neq 0} \langle [\hat{\bf N}_s]^2 \rangle + 3 ~,
\end{align}
namely, Eq.~\eqref{EW_nem}.

\paragraph{Generalization to arbitrary even-level systems.} A possible generalization to arbitrary integer spin is obtained by taking the projector to $\sum_{s=1}^{j}(\hat{n}_{a,s}-\hat{n}_{a,-s})$ for $a\in \{x,y,z\}$. The corresponding bound $\beta$ for local measurements $\{\{ \hat{n}_{a,s}\}_{s=-j\neq 0}^{j}\}_{a\in\{x,y,z\}}$ obtained numerically is summarized in the following table for several spin-$j$.  

 \begin{table}[h]
 \centering
	\begin{tabular}{| c | c | c | c | c | c |}
	\hline
	  spin $j$ & 1 & 2 & 3 & 4 & 5\\
	  \hline
	 $\beta$ for $\{\hat{n} \setminus \hat{\mathbf{n}}_0 \}$ & 1.25 & 1.5 & 1.5 & 1.3369 & 1.2926  \\
	 \hline
	 \end{tabular}
	\caption{Numerical separable bound $\beta$ for the basis of local measurements $\{\{ \hat{n}_{a,s}\}_{s=-j\neq 0}^{j}\}_{a\in\{x,y,z\}}$ for different spin-$j$.   }
	\label{table_bound_1}
	\end{table}
Notice that for $j=1$, we recover Eq.~\eqref{EW_singlet_spin-1}.

\section{Practical implementation of the algorithm}
\label{app_implementation}
In this appendix, we review step by step the implementation of the presented algorithm in order to reveal quantum entanglement from experimental data. The procedure consists of two parts. (i) First, we infer the required mean values from some hypothetical experiment. Then (ii), we apply our central result Ineq.~\eqref{eq_central_ineq} to (hopefully) detect entanglement. Note that our approach is system-agnostic, in the sense that it does not depend on features of the system not captured by the correlations considered. However, for the sake of concreteness, we will illustrate the implementation of the method with a concrete example. Consistently with the applications presented in the main text, we will consider a BEC of spin-1 atoms.

\subsection{Setting up the scenario}
The first step is to specify the partitioning of the considered quantum system to define entangled states. Specifically, here we consider an ensemble of $N$ three-level atoms; and we define a state as entangled if it cannot be decomposed as a mixture of product states over single-atom wave-functions.\\
The next step is to specify the single-atom observables $\{\{\hat{o}_k(i)\}_{k=1}^{K}\}_{i=1}^N$. Entanglement will be detected from correlations among them, so it must contain at least a pair of non-commuting observables. Furthermore, we consider the same set of local measurements for each atom. This simplification will allow us to infer the correlations in a scalable way from moments of collective operators, which is also the natural way in which measurements are performed in many cold-atom experiments. Concretely, in the present paper we consider two classes of local operators: (i) spin observables; and (ii) projectors to Zeeman sublevels. Here we work with $\{\{\hat{n}_{a,+1}(i), \hat{n}_{a,-1}(i) \}_{a=\{x,y,z \}}^{K}\}_{i=1}^N$, where $\hat{n}_{a,s}(i)$ is the projector to the hyperfine level $s$ along direction $a$ for atom $i$ as defined in the main text and $\{x,y,z\}$ an orthonormal basis of $\mathbb{R}^3$. Using a certain data to be specified in the next subsection, this setting leads to the witness Ineq. \eqref{EW_singlet_spin-1} of the main text, which we originally derived from a different set of local observables, namely for a combination of spin and populations to the zero state $\{\{\hat{s}_{a}(i), \hat{n}_{a,0}(i) \}_{a=\{x,y,z \}}^{K}\}_{i=1}^N$.

\subsection{Inferring the necessary statistics from the experiment}

According to the discussion of the main text, we need to compute the following expectation values:

\begin{equation}
    \left\{ \mathbf{m}_a = \frac{1}{N}(\langle \hat{N}_{a,+1}\rangle, \langle \hat{N}_{a,-1}\rangle), \  C_a = \frac{1}{N(N-1)}\begin{pmatrix}
        \langle \hat{N}_{a,+1} (\hat{N}_{a,+1}-1)\rangle & \langle \hat{N}_{a,+1} \hat{N}_{a,-1}\rangle \\
    \langle \hat{N}_{a,-1} \hat{N}_{a,+1}\rangle &  \langle \hat{N}_{a,-1} (\hat{N}_{a,-1}-1)\rangle
    \end{pmatrix}\right\}_{a\in\{x,y,z\}} \ ,
    \label{eq_MatrixStats}
\end{equation}
where $\hat{N}_{a,s}=\sum_{i=1}^N \hat{n}_{a,s}(i)$, that is, the population on hyperfine level $s$ along direction $a$. 

In our numerical illustration, we can consider $N= 100$ atoms in the ground state of the Hamiltonian of Eq. \eqref{H_spin-1} for $c = q = 1$. We compute the expectation values of Eq. \eqref{eq_MatrixStats} following the procedure detailed in Appendix \ref{Appendix_quantumdata}, which yields:
\begin{equation}
    \mathbf{m}_x = \mathbf{m}_y = (0.499,0.499), \mathbf{m}_z = (0.002,0.002)\ , C_x = C_y = \begin{pmatrix}
     0.248 & 0.251\\
     0.251 & 0.248
        \end{pmatrix}\ , C_z = \begin{pmatrix}
            0.000 & 0.000 \\
            0.000 & 0.000
        \end{pmatrix} \ .
    \label{eq_numbersMC}
\end{equation}

One can prepare such state on an unpolarized spin-1 BEC adiabatically by ramping down the magnetic field to $\sim \sqrt{q}$ from $q/c \gg 1$. Then, one needs to perform a collective Stern-Gerlach splitting along three orthogonal spatial directions, as achieved e.g. in ref.~\cite{evrardetal_science2021}. The statistics of \eqref{eq_MatrixStats} can readily be inferred by counting the number of atoms on each magnetic sublevel via fluorescence imaging \cite{evrardetal_science2021}.

\subsection{Detecting quantum entanglement}

Now we obtain the  matrices $\{C_a - \mathbf{m}_a\otimes \mathbf{m}_a := \tilde{C}_a\}_{a\in \{x,y,z\}} $, which are central to our method. With the values in Eq. \eqref{eq_numbersMC}, we have:  

\begin{equation}
    \tilde{C}_x = \tilde{C}_y = \begin{pmatrix}
        -0.139 & 0.140 \\
        0.140 & -0.139
    \end{pmatrix} \ ,
    \tilde{C}_z = \begin{pmatrix}
        0.000 & 0.002 \\
        0.002 & 0.000 
    \end{pmatrix} \ .
    \label{eq_Ctnumeric}
\end{equation}

Next, we diagonalize the $\tilde{C}$-matrices to compute the quantity $\mathcal{W} =\sum_{a\in \{x,y,z\}}(\sum_{\lambda < 0}\lambda(\tilde{C}_a) - \mathrm{Tr}(C_a)) $, where $\lambda(\tilde{C}_a)$ are the eigenvalues of $\tilde{C}_a$. In this example, it yields $\mathcal{W} =-1.550$. Comparing this value with the bound $\beta$ for the above set of local measurements, $\beta = 1.25$ (see Table \ref{table_bound_1} in Appendix \ref{app_EW_1}), we have $\mathcal{W} + \beta = -0.300<0 $. Therefore, according to the main inequality Eq. \eqref{eq_central_ineq} entanglement is detected. That is, there is no separable state compatible with the statistics contained in $\{C_a, m_a \}_{a\in\{x,y,,z\}}$. 

For the present case Eq. \eqref{eq_numbersMC}, the projectors to the negative spectrum are:
\begin{equation}
    P_x = P_y = P_z = \frac{1}{2}\begin{pmatrix}
        1 & -1 \\
        -1 & 1 
    \end{pmatrix} \ .
\end{equation}

Remarkably, we find the same optimal projectors regardless the value of the quadratic Zeeman strength $q$. Finally, from the set of local measurements, the projectors $\{P_x,P_y,P_z \}$ and $\beta$ one can derive the entanglement for all $N$ in terms of collective spin and population to the zero Zeeman sublevel as done e.g. in Appendix \ref{app_EW}. Once the entanglement witness Eq. \eqref{EW_singlet_spin-1} is derived, its violation can be verified equivalently from the expectation values $\langle \hat{\mathbf{J}}^2 \rangle, \langle \hat{\mathbf{J}} \rangle, \langle \hat{\mathbf{N}}_0^2 \rangle $.


\end{document}